\documentclass{article}
\usepackage{arxiv}
\usepackage[utf8]{inputenc}
\usepackage[T1]{fontenc}
\usepackage{amsmath,amssymb,amsfonts}
\usepackage{graphicx}
\usepackage{booktabs}
\usepackage{siunitx}
\usepackage{float}
\usepackage[title]{appendix}
\usepackage{subcaption}
\usepackage{placeins}
\usepackage{microtype}
\usepackage{etoolbox}
\usepackage[numbers,sort&compress]{natbib}
\usepackage{xurl}
\usepackage{hyperref}
\usepackage{doi}

\renewcommand{\shorttitle}{Keystroke dynamics and temporal flexibility}
\makeatletter
\patchcmd{\@maketitle}
  {\vskip 0.4in \@minus 0.1in \center{\@date} \vskip 0.2in}
  {\vskip 0.18in}{}{\PackageError{paper2}{Title-spacing patch failed}{Check arxiv.sty.}}
\makeatother
\title{Combining heavy-tailed keystroke dynamics with eye tracking reveals temporal flexibility in human problem solving}

\author{%
\href{mailto:dinc.yavuz@lmu.de}{Yavuz Dinc}\textsuperscript{1,*}\quad
\href{mailto:f.kieser@fu-berlin.de}{Fabian Kieser}\textsuperscript{2}\quad
\href{mailto:peter.wulff@ph-ludwigsburg.de}{Peter Wulff}\textsuperscript{3}\quad
\href{mailto:johannes.zierenberg@ds.mpg.de}{Johannes Zierenberg}\textsuperscript{4,5}\\[4pt]
\bfseries\href{mailto:Niklas.Stausberg@lmu.de}{Niklas Stausberg}\textsuperscript{1}\quad
\href{mailto:K.Avila@physik.uni-muenchen.de}{Karina E. Avila}\textsuperscript{1}\quad
\href{mailto:Jochen.Kuhn@lmu.de}{Jochen Kuhn}\textsuperscript{1}\quad
\href{mailto:s.kuechemann@lmu.de}{Stefan K\"uchemann}\textsuperscript{1}\\[8pt]
\normalfont\small\textsuperscript{1}Ludwig-Maximilians-Universit\"at M\"unchen, Germany\\
\normalfont\small\textsuperscript{2}Freie Universit\"at Berlin, Germany\\
\normalfont\small\textsuperscript{3}P\"adagogische Hochschule Ludwigsburg, Germany\\
\normalfont\small\textsuperscript{4}Max-Planck-Institut f\"ur Dynamik und Selbstorganisation, G\"ottingen, Germany\\
\normalfont\small\textsuperscript{5}Institute for the Dynamics of Complex Systems, University of G\"ottingen, Germany\\[5pt]
\normalfont\small\textsuperscript{*}Correspondence: \href{mailto:dinc.yavuz@lmu.de}{\texttt{dinc.yavuz@lmu.de}}
}

\hypersetup{
  pdftitle={Combining heavy-tailed keystroke dynamics with eye tracking reveals temporal flexibility in human problem solving},
  pdfauthor={Yavuz Dinc, Fabian Kieser, Peter Wulff, Johannes Zierenberg, Niklas Stausberg, Karina E. Avila, Jochen Kuhn, Stefan Kuechemann},
  pdfsubject={Temporal flexibility in human problem solving},
  pdfkeywords={keystroke dynamics, eye tracking, heavy-tailed distributions, problem solving}
}

\renewcommand{\undertitle}{}  
\renewcommand{\headeright}{} 

\begin{document}
\maketitle

\begin{abstract}
Human problem solving unfolds through complex temporal dynamics that reflect the coordination of perception, cognition, and action. The fine structure of this process is still not very well understood. We use high resolution keystroke logging combined with synchronized eye tracking to examine the micro-temporal structure of human problem solving during written solution construction. Looking at 131{,}718 inter-keystroke intervals,
we find heavy-tailed timing patterns spanning multiple orders of magnitude within the observed range. A three-component Gaussian mixture summarizes overlapping short, intermediate, and long pause ranges. Triangulation with gaze data reveals a systematic shift of attention toward external representations with increasing pause duration. Associations between tail exponents and performance differ across tasks. A larger across-task range of tail exponents ($\alpha_{\mathrm{range}}$) is positively associated with overall performance after adjusting for prior knowledge and measured process covariates ($\beta = 0.351$, $p_{\mathrm{HC3}} = 0.0165$), accounting for an additional 11.3 percentage points of explained variance in this sample. These findings are consistent with the hypothesis that successful reasoning involves adaptive reconfiguration of temporal organization across task demands. The heavy-tailed timing dynamics represent a signature of task dependent cognitive organisation and highlight the potential of keystroke data as an unobtrusive and scalable window into the process of problem solving.
\end{abstract}


\section{Introduction}\label{sec1}

Complex cognitive tasks (e.g. explaining a scientific concept in writing, solving a multi-step problem, writing code) don't proceed as a continuous stream of activity but as temporally structured episodes of production and pause \cite{Alves2008,Shrestha2022}. They are typically characterised by periods of rapid production interspersed with pauses that can range from a few tens of milliseconds to many seconds. This intermittency is not just noise: across cognition and behaviour there is a tendency for temporal patterns to be heavy-tailed and self-similar, with $1/f$-like fluctuations and bursty inter-event times \cite{gilden2001cognitive,vanorden2003selforg,kello2010scaling}. Work on human activity records further show bursty temporal organization and heavy-tailed inter-event time distributions \cite{Barabasi2005,Vazquez2006,Goh2008, ShengEtAl2023, IacopiniEtAl2024}. The distribution of interval durations and
dependencies in their temporal ordering are distinct properties. These findings motivate a
distributional analysis of pause timing, but do not by themselves establish an exact power-law
mechanism \cite{Clauset2009,Newman2005}. A related measurement challenge is that most studies of complex reasoning still rely on static outcomes (final answers) or coarse timing measures (total time-on-task). These summaries collapse the within-task dynamics that likely matter for success: planning and monitoring, updating partial solutions, switching strategies, and integrating external information \cite{Goldhammer2014TimeOnTask,Stadler2023InteractiveTasksLSA,Eichmann2019PlanningCPS}. In other words, what distinguishes successful from unsuccessful solvers may not be a single ``speed'' parameter, but how flexibly they reconfigure processing modes as task demands change.

Process-tracing methods offer a way to study these dynamics directly. In writing research,
keystroke logging has long been used to capture fine-grained production traces and to operationalize
pauses and bursts as windows into ongoing cognitive processes
\cite{HayesFlower1980,Kellogg1996,Matsuhashi1981,Schilperoord1996,Wengelin2006,Leijten2013,Baaijen2012}.
Pauses are commonly interpreted as reflecting transitions between nested subprocesses---motor
execution, lexical/formulation processes, and higher-level planning or revision---yet pause
distributions are typically skewed and heterogeneous, making threshold-based analyses and mean-based
summaries fragile \cite{Wengelin2006,Baaijen2012}. Recent work therefore emphasizes distributional
approaches and mixture perspectives to describe multiple, potentially overlapping time scales in the
same observable trace \cite{Baaijen2012,li2021identifying,hall2024constructing}. Keystroke
time-series studies have also linked fractal scaling to essay quality beyond essay length
\cite{likens2017keystroke}. These time-series measures are distinct from the tail exponents of
interval distributions studied here. Timing data alone, however, are often underspecified: the same pause duration can correspond to qualitatively different cognitive states. A principled way to increase interpretability is to triangulate production traces with measures of information uptake. Eye tracking provides a complementary view by indexing how attention is allocated across text, diagrams, and other external representations, and by revealing gaze patterns related to information uptake and comprehension \cite{JustCarpenter1980,Henderson2013,alemdag2018systematic, MurlidaranEckstein2026}. 

In multimedia and problem solving contexts gaze becomes more selective and task relevant with expertise in line with information reduction accounts and goal directed control \cite{haider1999eye,Jarodzka2010EyesOfBeholder}. Combining eye tracking with keystroke logging is therefore particularly promising in that it can link the micro temporal structure of pauses to moments of perceptual sampling, representational integration and strategy revision \cite{Wengelin2009}.

Beyond the interpretation of individual pauses there is a second question to be asked, namely whether the temporal organisation of activity changes as a function of task demands. In this context, adaptive expertise distinguishes efficient execution of familiar
procedures from the capacity to reorganize processing when routine procedures are insufficient
\cite{schwartz2000efficiency,hatano1984two}. This does not necessarily mean that the successful solver is always going to be the faster one. Rather, it motivates the expectation that the
balance between fluent production and extended episodes of information search, monitoring, or
reformulation may vary across tasks. Work on external representations has also demonstrated that adaptation can involve an initial investment of time in reorganising information for future use
\cite{martin2009prospective}. Crucially, adaptive strategy use is about the match between behaviour and the task, the person and the context rather than simply the amount of variation
 \cite{verschaffel2009conceptualizing, PoliEtAl2024}. People can differ in their degree of adjustment of strategy use as a function of changing rates of success and this can in turn be relevant to performance \cite{schunn2001another}. We therefore make a distinction between task-specific associations between timing and performance and the amount of temporal variation across tasks in the same individual. The latter may be consistent with temporal flexibility but does not establish that the
variation is appropriate to each task. We use this framework to motivate tests of task-contingent
temporal organization rather than to operationalize adaptive expertise directly.

In the present work we study these dynamics in a setting that naturally elicits switches between internal formulation and external information integration: multi-representational quantitative reasoning problems. These tasks require the construction of a situational model, the mapping of that model onto formal representations and the continual coordination of text, diagrams and symbolic representations \cite{larkin1980expert,reif1982knowledge,heller1984prescribing,ainsworth2006deft}. Performance differences are often explained in terms of representational competence and strategic control—how solvers select, translate, and integrate representations, and how they change approach when progress stalls \cite{kohl2005representational,tuminaro2007epistemic}. The demands of this task class make it an ideal setting in which to explore principles of task-contingent temporal organisation during complex cognition. We combine high resolution keystroke logging with concurrent eye tracking to characterise the temporal organisation of written solution construction. Using data from
university students solving six open-ended, multi-representational physics problems, we ask
(i) whether inter-keystroke intervals exhibit heavy-tailed timing, (ii) whether multiple,
potentially overlapping time scales can be characterized within the pause distribution, and
(iii) how these time scales map onto gaze behavior during pauses. Building on accounts of
cognitive flexibility and task-contingent control
\cite{miyake2000unity,monsell2003task,dajani2015demystifying}, we then test whether the
association between estimated tail exponents and performance differs across tasks. We finally examine, in a person-level analysis, whether between-task
variation is associated with overall success beyond prior knowledge
and measured keystroke covariates.
Complementary analyses look at the performance associations of the mixture parameters. Together, this approach examines fine-grained temporal organization
and its associations with visual sampling and performance, with
possible implications for data-driven assessment and adaptive
support in cognitive tasks.

\section{Results}\label{sec2}
\subsection{\label{sec:subsection-1}Keystroke Analysis}

\begin{figure}[t]
  \centering
  \includegraphics[width=0.9\textwidth]{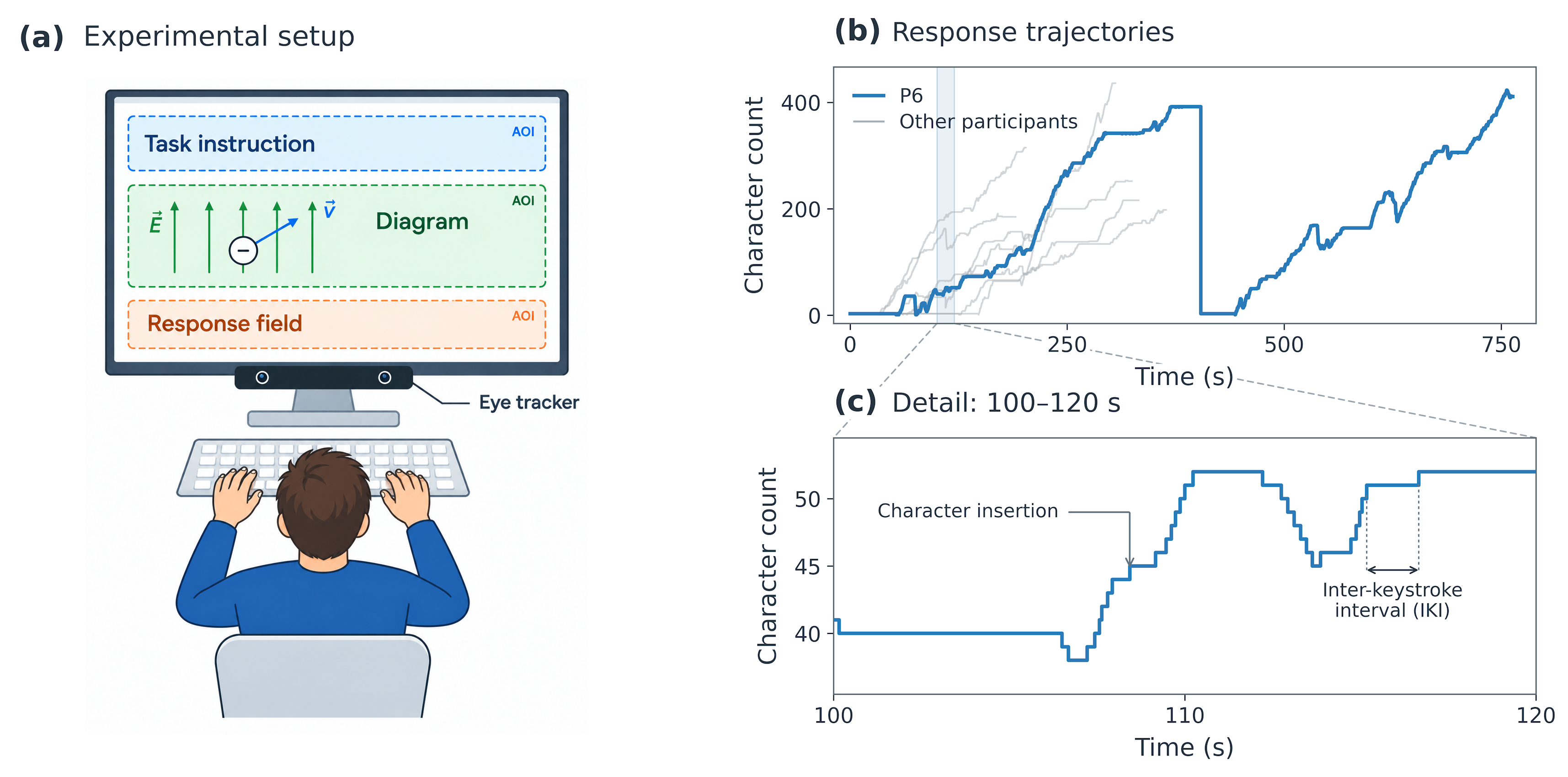}
\caption{\label{fig:keyovertime}
    \textbf{(a)}~Schematic of the experimental setup: participants 
    solved open-ended physics problems on screen with their gaze 
    recorded by an eye tracker mounted below the display. 
    The dashed rectangles indicate the areas of interest (AOIs) used 
    for gaze analysis.
    \textbf{(b)}~Character count over time for all 
    participants on one item. The grey traces are the other participants, 
    the highlighted trace (P6) shows a full response revision 
    occurring around $t \approx 400$\,s. The shaded region is the time 
    window shown in~(c).
    \textbf{(c)}~Magnified 20-second window showing the discrete, 
    stepwise structure of individual keystrokes: the rising segments 
    are character insertions, the flat segments are IKIs, and the downward 
    steps are deletions.}
\end{figure}

For the pause time analysis we calculated inter-keystroke intervals (IKIs), defined as the time between two consecutive keystrokes in the same input field (i.e. writing pauses). The analysis was restricted to the active writing period (so the pre-typing interval i.e. time spent reading the prompt was excluded). For each participant and each open response item this gives us a typing trajectory (character count over time) with the initial latency being the time taken to make the first input, the local slope being the instantaneous typing speed and the flat sections being the pauses (including the micro-pauses which may not be resolved at the at plot scale; see Fig.~\ref{fig:keyovertime}). We use the IKIs to quantify pause durations \(t_{\mathrm{pause}}\) and characterize their distribution.
As shown in Fig.~\ref{fig:keyovertime}, during the active writing phase, the writing trajectories contain roughly linear segments interspersed with IKIs. One participant (P6) revises the entire response at a certain point and begins a new approach.

\subsection{Pause-time distributions}

\begin{figure}[tbp]
  \centering

  \begin{subfigure}[b]{0.49\linewidth}
    \centering
    \includegraphics[width=\linewidth]{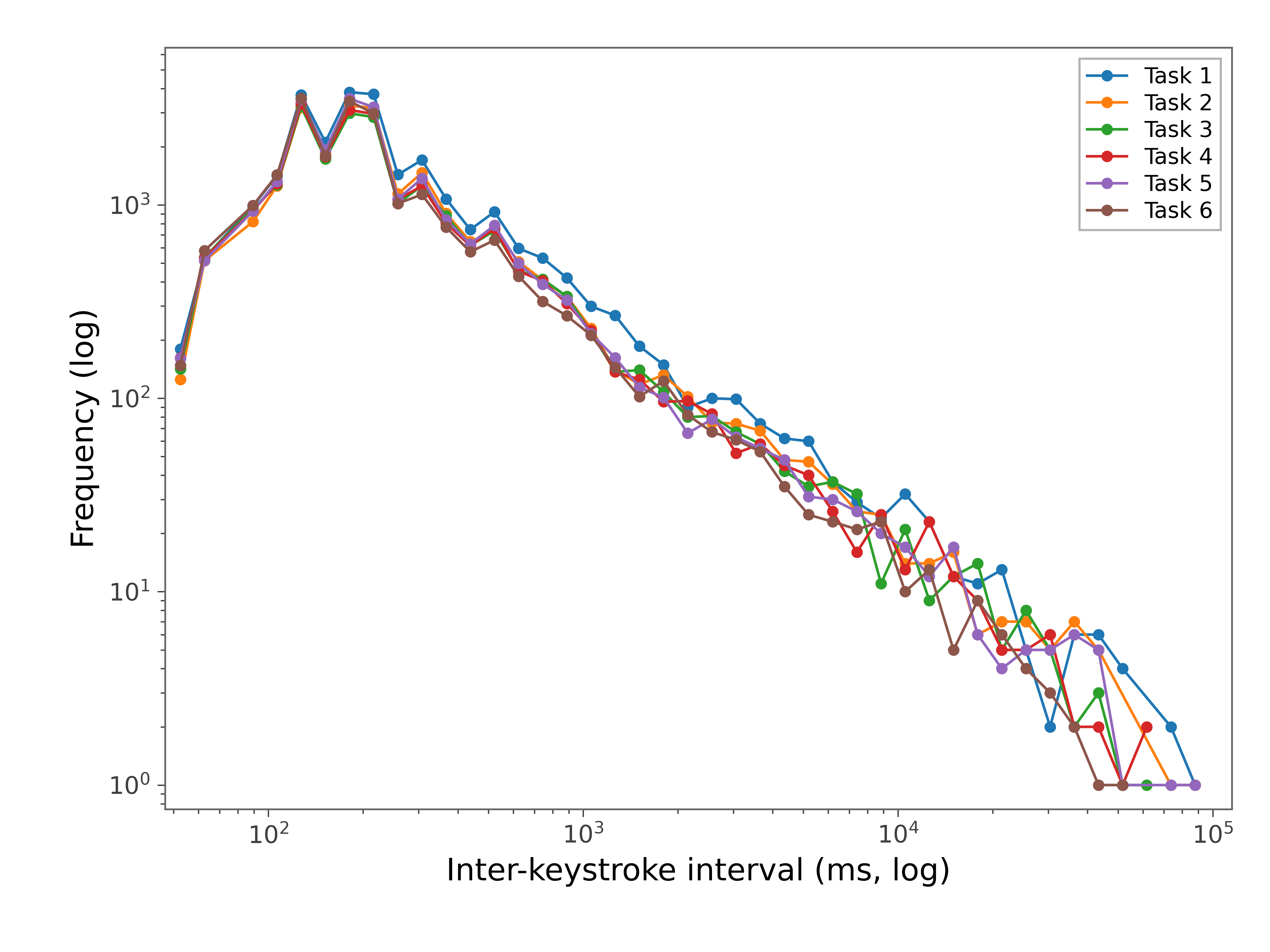}
    \caption{By task}
    \label{fig:pause-by-task}
  \end{subfigure}
  \hfill
  \begin{subfigure}[b]{0.49\linewidth}
    \centering
    \includegraphics[width=\linewidth]{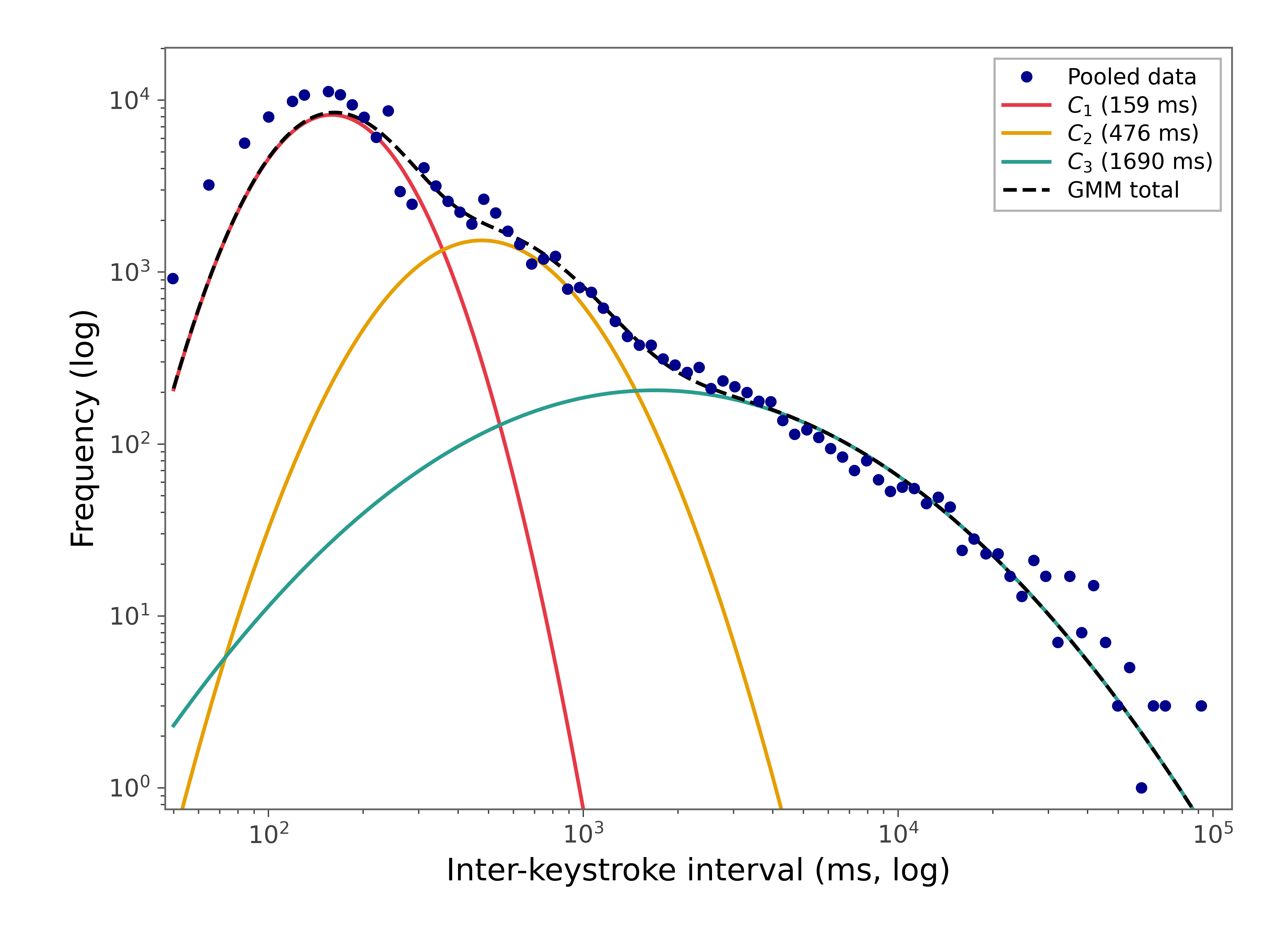}
    \caption{Pooled}
    \label{fig:pause-all}
  \end{subfigure}

  \caption{Log--log histograms of IKIs plotted for
  \textbf{(a)} individual tasks and \textbf{(b)} pooled data across all tasks.
  The distributions show similar shapes with upper-tail curvature.
  In \textbf{(b)}, points show empirical bin counts; solid colored lines show
  the component contributions of a three-component Gaussian mixture model
  fitted to \(\log_{10}\)-transformed IKIs, and the dashed black line shows
  their sum. Values in the legend denote component medians on the original
  time scale.}
  \label{fig:pause-loglog}
\end{figure}

From the raw logs, we extracted IKIs, obtaining
\(131{,}718\) intervals from \(\mathrm{N = 55}\) participants after restricting
the analysis to \(50\,\mathrm{ms} \leq t_{\mathrm{pause}} \leq 100\,\mathrm{s}\).
Figure~\ref{fig:pause-by-task} shows log--log histograms of IKIs by task; the
pooled distribution is shown in Fig.~\ref{fig:pause-all}. The complementary cumulative distribution function (CCDF),
$S(t)=P(T\geq t)$, gives the probability that an inter-keystroke
interval $T$ lasts at least $t$. Empirically, it is the proportion
of observed intervals with duration at least $t$.
The CCDF of the pooled IKIs is shown in
Fig.~\ref{fig:pause-ccdf}.
Participant-level histograms and CCDFs are provided in
Appendix Fig.~\ref{fig:supp-participant-distributions}. The
curves are very similar and essentially parallel across tasks, with their peak frequencies
at around 100--200\,ms. The tail (i.e. the upper tail corresponding to longer pauses) is broadly linear on log–log axes over a fairly limited range, as would be expected for a long right tail. The upper tail, meaning the right-hand part of the distribution
corresponding to longer pauses, bends below a linear extrapolation and
becomes increasingly sparse at long intervals. The similarity of the task specific curves does lend support to a pooled description of the overall distribution and we therefore go on to analyse the pooled distribution in the next section.

\begin{figure}[tbp]
  \centering
  \includegraphics[width=0.8\linewidth]{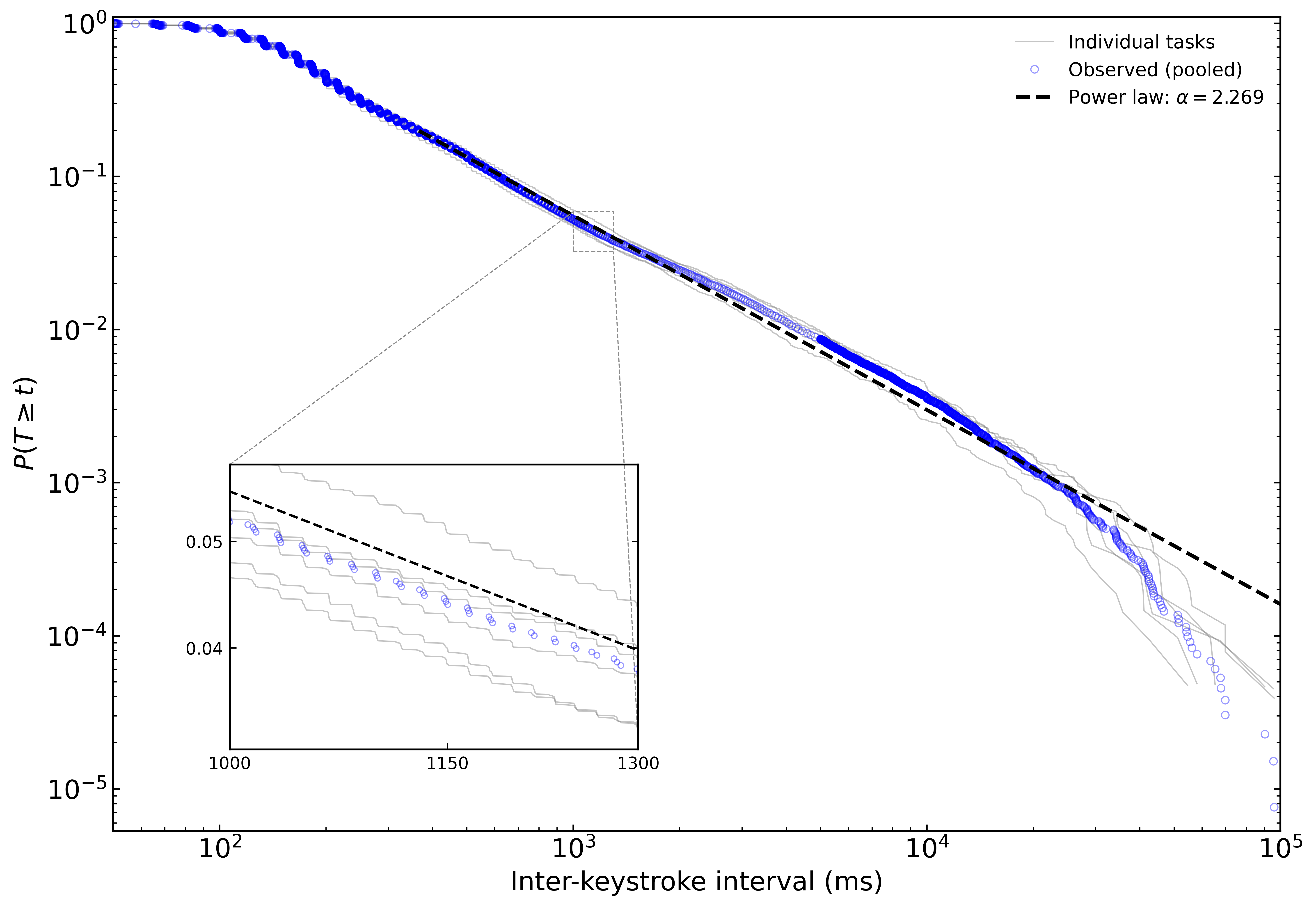}
 \caption{\label{fig:pause-ccdf} Complementary cumulative distribution functions of IKIs within the
50-ms--100-s analysis window, pooled across tasks (blue circles)
and shown separately for each task (grey lines).
The dashed line shows the reference power-law fit
($\hat{\alpha}=2.2692$), estimated from
$n_{\mathrm{tail}}=26{,}168$ intervals between
$x_{\min}=366.183$\,ms and 100\,s. The
  plot highlights an approximately linear regime over part of the observed
  range, followed by increasing curvature and sparsity in the far right tail.}
\end{figure}

\subsection{\label{sec:subsection-2}Power Law}

We examined whether the upper tail of the IKI distribution could be described by a power law, using maximum-likelihood estimation and Kolmogorov--Smirnov (KS) diagnostics~\cite{Clauset2009}. For the pooled descriptive fit, we chose the local KS minimum corresponding to the lower cutoff, $x_{\min}=366.183$\,ms. The alternative KS minimum produced only a very slight change in the estimated exponent. The corresponding reference fit, using a density with unbounded upper support, yielded $\hat{\alpha}=2.2692$. The pooled power-law fit used $n_{\mathrm{tail}}=26{,}168$ IKIs satisfying $366.183$\,ms $\leq t_{\mathrm{pause}}\leq100$\,s. Here, $n_{\mathrm{tail}}$ denotes the number of IKIs in the fitted upper-tail range. These IKIs constitute $19.87\%$ of the $131{,}718$ IKIs retained in the overall 50-ms--100-s analysis window across 55 participants. On log--log axes, the CCDF shows an approximately linear region with upper-tail deviations. Within the fitted tail, lower $\alpha$ indicates a slower decay and a greater relative contribution of longer pauses, whereas higher $\alpha$ indicates a stronger concentration near $x_{\min}$.

For model comparison all of the candidate densities were conditioned on the same fitted interval with the selected lower cut-off held fixed. This produced a very similar power law exponent. The lognormal alternative reached the power law boundary and provided no improvement in maximised log likelihood. The stretched-exponential and exponentially truncated power-law alternatives likewise approached the power-law boundary, whereas the exponential model fitted substantially worse. Separate parametric-bootstrap KS diagnostics indicated lack of fit of the shared power-law/lognormal limiting density. These diagnostics assume independent and identically distributed intervals and hold the cutoff fixed; they do not account for within-participant dependence or threshold selection. We therefore treat the power-law fit as a descriptive approximation rather than an exact generative model. Full model comparisons and further diagnostics are given in Appendix~\ref{app:tail-diagnostics}. As a robustness check we also considered the reference exponent obtained when restricting to intervals between consecutive single character insertions and this gave a very similar value. The exponent was therefore similar when deletion-related intervals were excluded. As a further descriptive check, reference fits with separately KS-selected cutoffs yielded estimates for all 55 participants (pooling across tasks) and all six tasks (pooling across participants). Participant-level exponents were centered around the pooled estimate ($\alpha$: $M=2.31$, $SD=0.19$). Task-level exponents were similar across the six problems ($M=2.25$, $SD=0.031$), consistent with the visual overlap of task-specific IKI distributions. These descriptive estimates do not in themselves constitute separate goodness-of-fit confirmations at either level. Collectively the results do support a parsimonious power-law-like description of the tail that we observe, without establishing scale invariance or a unique generative distribution. We go on in the following sections to supplement this tail description with a Gaussian mixture model (GMM) of the log-IKI distribution and within-pause gaze analyses.

\subsection{Mixture structure of pause times (GMM)}

We asked whether the empirical pause distribution could be described by
multiple overlapping time scales rather than a single component. To this
end, we modeled $\log_{10}(t_{\mathrm{pause}}/\mathrm{s})$ with Gaussian
mixture models (GMMs) and compared model orders $K=1,\ldots,9$ using AIC
and BIC (Fig.~\ref{fig:gmm_bic_aic}). Although $K=4$ improved both criteria
in the pooled data, a matched comparison of 320 participant--task cells
favored $K=3$ over $K=4$ by BIC in 315 cells. We therefore retained $K=3$
as a parsimonious common parameterization. Fitting a three-component GMM to $131{,}718$ IKIs between 50\,ms and
100\,s yields overlapping short, intermediate, and long components
(Fig.~\ref{fig:pause-all}). On the original scale, component medians
($10^{\mu_k}$\,s) are $0.159$\,s (short), $0.476$\,s (intermediate), and
$1.690$\,s (long), with weights of $0.761$, $0.186$, and $0.052$,
respectively (Table~\ref{tab:gmm-params}). The long-pause component
exhibits the largest dispersion on the $\log_{10}$ scale, indicating broad
variability in extended pauses. A tentative functional interpretation (consistent with previous accounts of the writing process) would be that the short component is associated with motor execution during fluent typing, the intermediate component with local/lexical operations and the long component with reflective planning/monitoring. These are all very tentative interpretations and are not established by the GMM alone. In the next section we relate these time scales to gaze behaviour within the pauses (AOI dwell time and fixations).

\subsection{\label{sec:subsection-3}Eye Tracking across Pause Durations}

We've described the three overlapping components of the pause time distribution using the GMM and now turn to where participants were looking during the pauses. Each open response item was systematically broken down into pre-defined Areas of Interest (AOIs) which included
the \emph{problem stem}, \emph{graph region}, \emph{response field},
and \emph{instructions} (see Appendix). Other defined regions and valid gaze outside of the defined AOIs were kept as a separate category. For each pause we mapped the observed gaze to the AOIs and then calculated (i) the proportion of pooled valid dwell time in each AOI within each pause category and (ii) the mean fixation duration within each category. For the purposes of visualisation, pauses in the 0.05--100\,s analysis window were
grouped into four ordered duration categories (Short, Medium,
Long, and Very Long), using GMM-derived boundaries of approximately
0.358, 1.470, and 16.902\,s. The Very Long category is thus isolating the upper tail of the longest component rather than representing a fourth fitted component. 

\begin{figure}[htbp]
  \centering

  \textbf{(a)}\par\vspace{1mm}
  \includegraphics[width=0.76\linewidth]{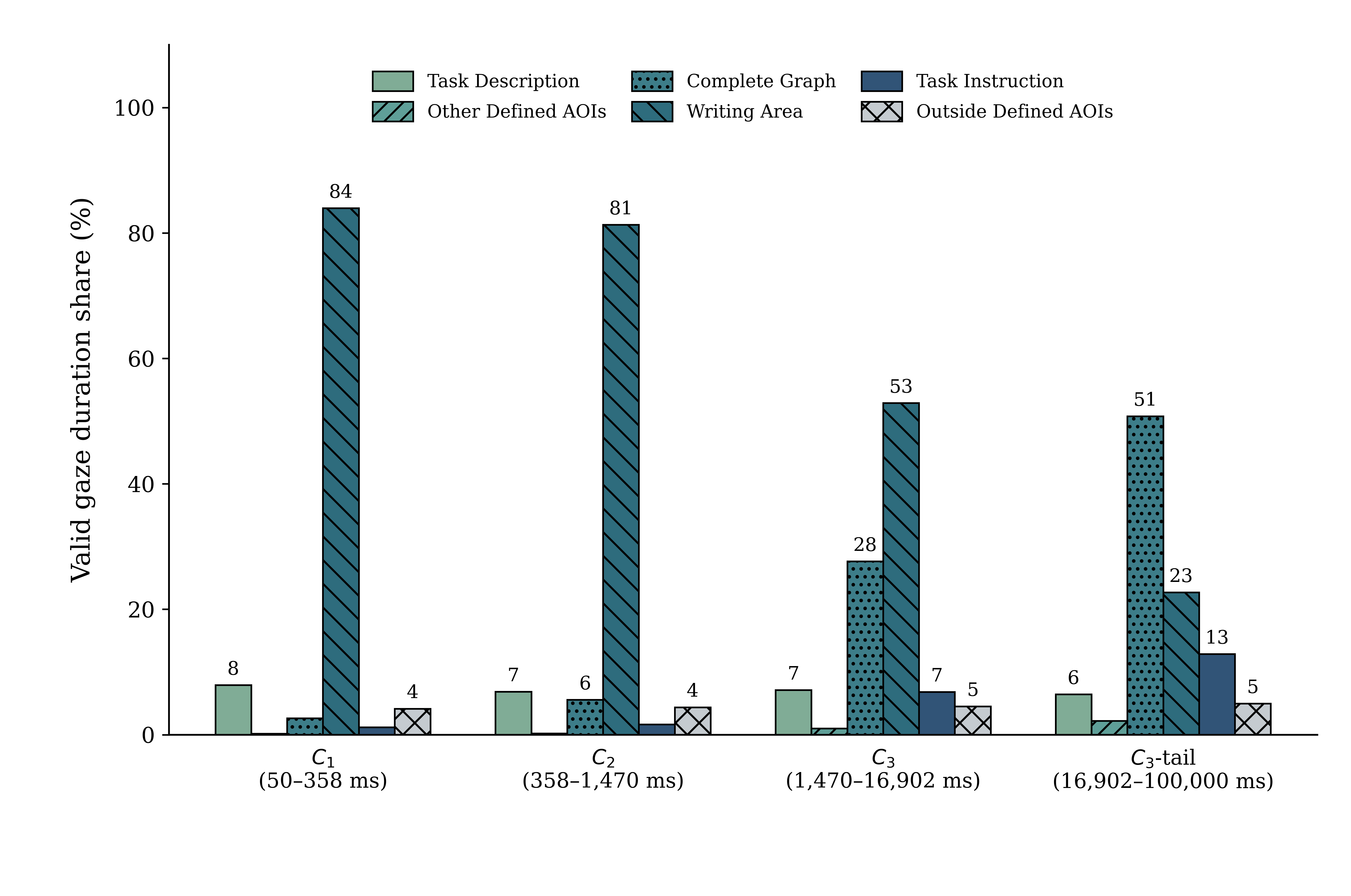}

  \vspace{3mm}

  \textbf{(b)}\par\vspace{1mm}
  \includegraphics[width=0.76\linewidth]{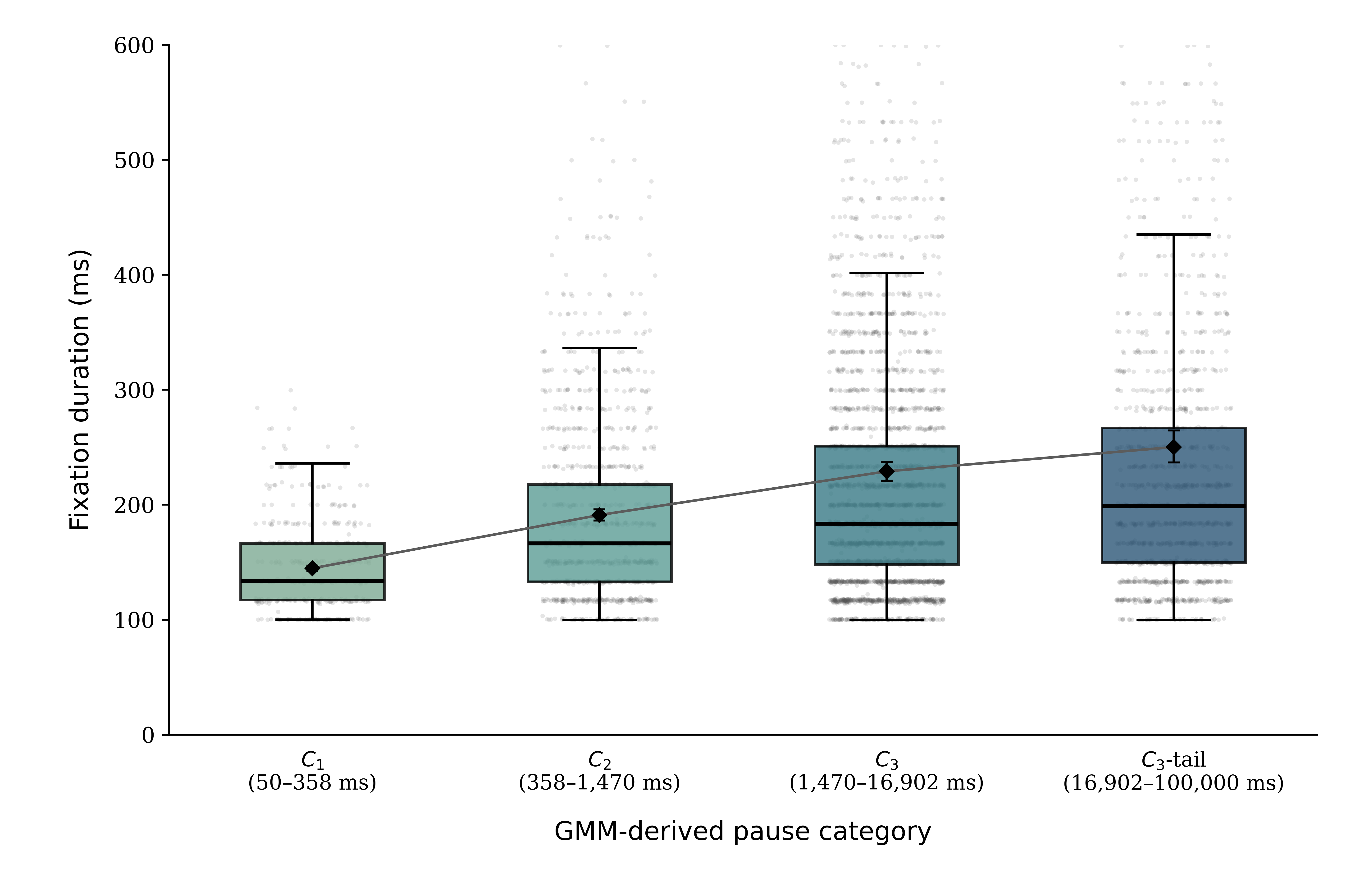}

  \caption{\label{fig:gaze_fixations}
  Eye-tracking measures across four GMM-derived pause duration categories.
  \textbf{(a)}, Proportion of pooled valid gaze dwell time per AOI,
  including valid gaze outside defined AOIs; missing gaze is excluded.
  \textbf{(b)}, Distributions of individual fixation durations.
    Boxes show medians and interquartile ranges; gray points show
    a 10\% sample of fixations. Black diamonds indicate equally
    weighted means of participant-level category means with
    95\% participant-bootstrap confidence intervals.
  }
\end{figure}

There is a clear shift in the observed gaze allocation as pause length increases (Fig.~\ref{fig:gaze_fixations}a). During
short pauses, gaze was concentrated in the \emph{response field}
(approximately 84\% of pooled valid dwell time), consistent with
attention to ongoing text production. As pauses became longer,
the share of dwell in the \emph{response field} decreased to
approximately 81\%, 53\%, and 23\%, while dwell in the
\emph{graph region} increased from approximately 3\% in short
pauses to 6\%, 28\%, and 51\%, respectively. In the longest
pauses, graph-directed dwell thus exceeded response-field dwell,
consistent with greater visual engagement with an external task
representation. Mean fixation duration increased monotonically with pause category
(Fig.~\ref{fig:gaze_fixations}b). To account for the nested data
structure, the 58{,}260 valid fixations fully contained within a
pause were averaged per participant~$\times$~pause category before
inferential testing. Subject-level means were 145, 191, 229, and
250\,ms for Short, Medium, Long, and Very Long pauses, respectively
(95\% participant-bootstrap CIs: 142--147, 186--196, 221--237,
and 237--265\,ms). Adjacent-category comparisons were positive
after Benjamini--Hochberg correction (all $p_{\mathrm{BH}} < .001$).
A supplementary random-intercept model fitted to the 214
participant~$\times$~category means yielded
$\mathrm{ICC} = .361$. Thus, longer pauses were associated with
longer fixation durations in participant-level summaries.

\subsection{\label{sec:subsection-4}Correlation with Students' Performance}

To investigate the relationship between the temporal dynamics of the writing process characterized by the fitted tail exponent $\alpha$ and the achieved performance in physics tasks, correlation analyses were conducted. First, we examined whether a general relationship exists between the exponent and the total performance across all six tasks. The analysis revealed no significant correlation between the global alpha value and the total score ($r = -0.130, p = .343$).

\begin{figure}[t]
  \centering
  \includegraphics[width=0.9\textwidth]{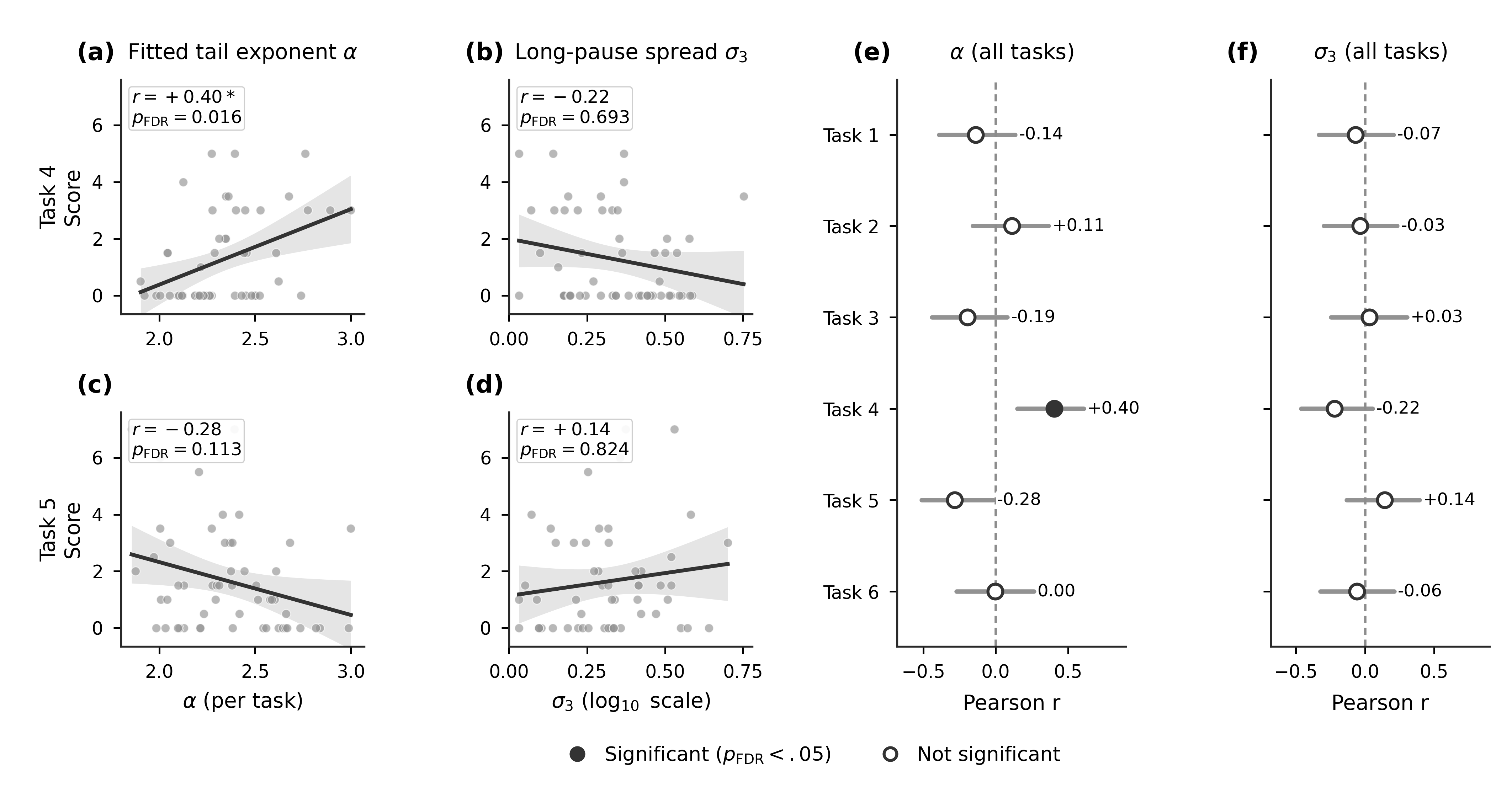}
  \caption{\label{fig:corr}Task-dependent timing--performance associations.
  (a--d)~Per-task associations between performance and two timing descriptors---the fitted tail
  exponent $\alpha$ (a,~c) and the long-pause component width $\sigma_3$ (b,~d)---for
  Task~4 and Task~5. Points: participants with available fits; lines: OLS fit
  $\pm$~95\% CI; insets: Pearson $r$ and FDR-adjusted $p$ (asterisk: $p_{\mathrm{FDR}}<.05$).
  (e,~f)~Task-level correlations with performance for $\alpha$ (e) and $\sigma_3$ (f) across
  all six tasks (whiskers: 95\% CI). Filled markers survive Benjamini--Hochberg FDR
  correction, applied separately across the six tasks for each descriptor; open markers
  do not; sign is given by position relative to zero. Only the positive $\alpha$--performance
  association in Task~4 survives FDR correction; none of the $\sigma_3$ associations do. The formal task-by-$\alpha$ interaction supports differences in the $\alpha$--performance association across tasks (likelihood-ratio $\chi^2(5) = 19.93$, $p = .0013$).}
\end{figure}

However, a task-level analysis revealed that the association between
pause dynamics and performance is context-dependent (Figure~\ref{fig:corr}).
The $\alpha$ analyses included 323 valid participant--task estimates from 55 students.
To account for six simultaneous comparisons, task-level $p$-values
were adjusted using the Benjamini--Hochberg procedure controlling the
false discovery rate at .05, separately for each timing descriptor. No statistically significant relationship was detected for Tasks~1, 2, 3, and~6
(all $p_{\mathrm{FDR}} \geq .328$). The estimates for Tasks~4 and~5 pointed in
opposing directions, but only Task~4 survived FDR correction:

\begin{itemize}
    \item \textbf{Task 4:} A significant positive correlation was observed
    ($r = 0.404$, $p_{\mathrm{FDR}} = .016$), indicating
    that higher $\alpha$ values were associated with better performance.

    \item \textbf{Task 5:} In contrast, the sample correlation was negative
    ($r = -0.284$, $p_{\mathrm{FDR}} = .113$), but did not reach statistical
    significance after FDR correction.
\end{itemize}

For the width of the long-pause component, $\sigma_3$, no task-specific
association with performance survived FDR correction
(all $p_{\mathrm{FDR}} \geq .693$; Figure~\ref{fig:corr}). A formal test of the task-by-$\alpha$ interaction supported differences in
this association across tasks (likelihood-ratio $\chi^2(5) = 19.93$,
asymptotic $p = .0013$). A participant fixed-effects analysis with
participant-clustered standard errors likewise supported the interaction
($p = .00042$).

In summary, these results suggest that the relationship between the temporal
organization of writing and performance varies across tasks, rather than
being captured by a single task-independent association. This pattern is
compatible with adaptive reconfiguration in response to task-specific
cognitive demands, but does not by itself establish such adaptation.

\subsection{\label{sec:temporal_adaptivity}Across-task variability and problem-solving success}

The task-dependent associations between $\alpha$ and performance
motivated an analysis of whether within-person variation in fitted
tail exponents across tasks was associated with overall problem-solving
success. We summarized this variation as $\alpha_{\mathrm{range}}$,
the difference between each student's maximum and minimum valid
task-specific exponent. A larger $\alpha_{\mathrm{range}}$ therefore reflects greater across-task variation in fitted tail decay.

\begin{figure}[t]
  \centering
  \begin{minipage}[t]{0.45\textwidth}
    \textbf{(a)}\par
    \includegraphics[width=\linewidth]{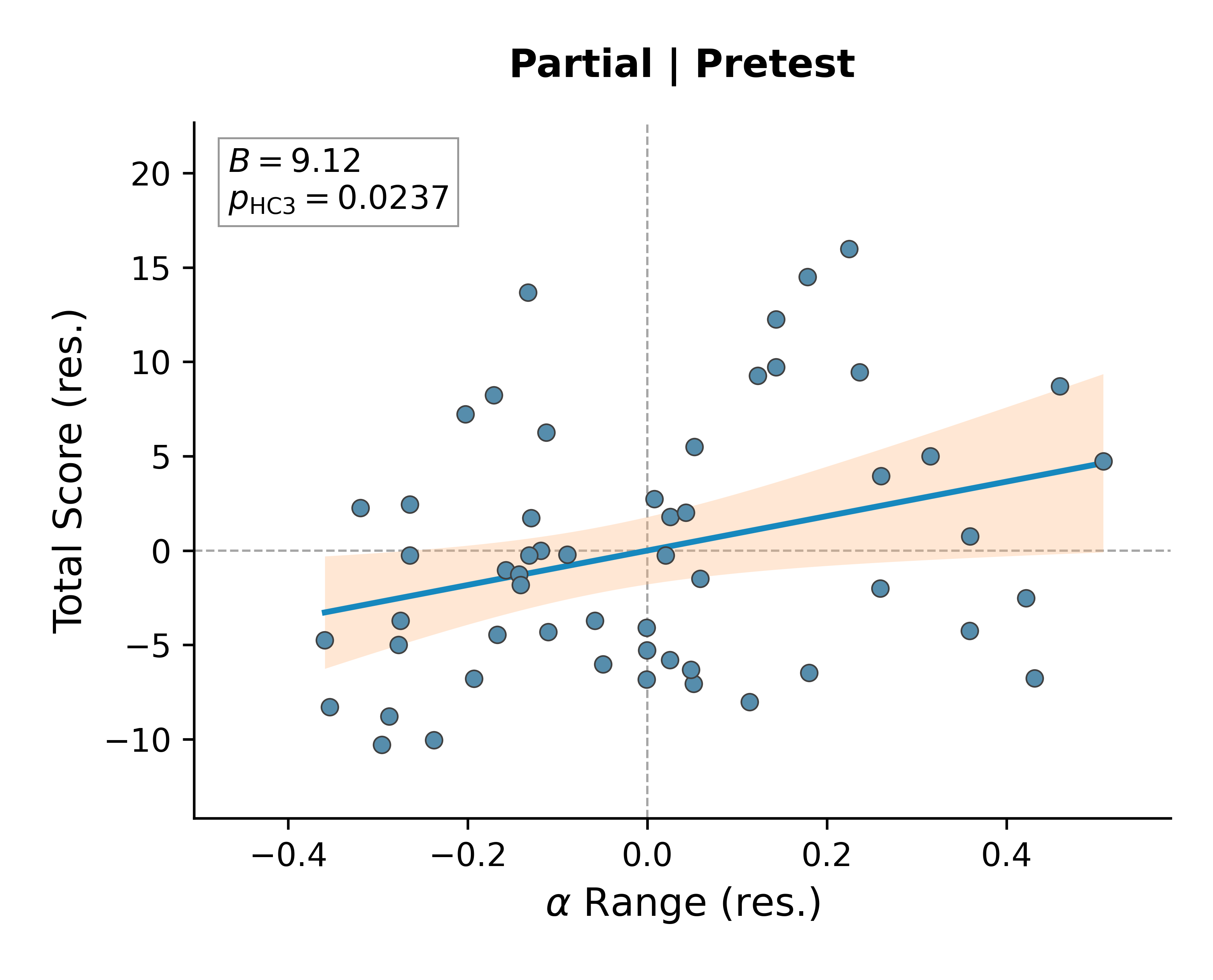}
  \end{minipage}\hfill
  \begin{minipage}[t]{0.45\textwidth}
    \textbf{(b)}\par
    \includegraphics[width=\linewidth]{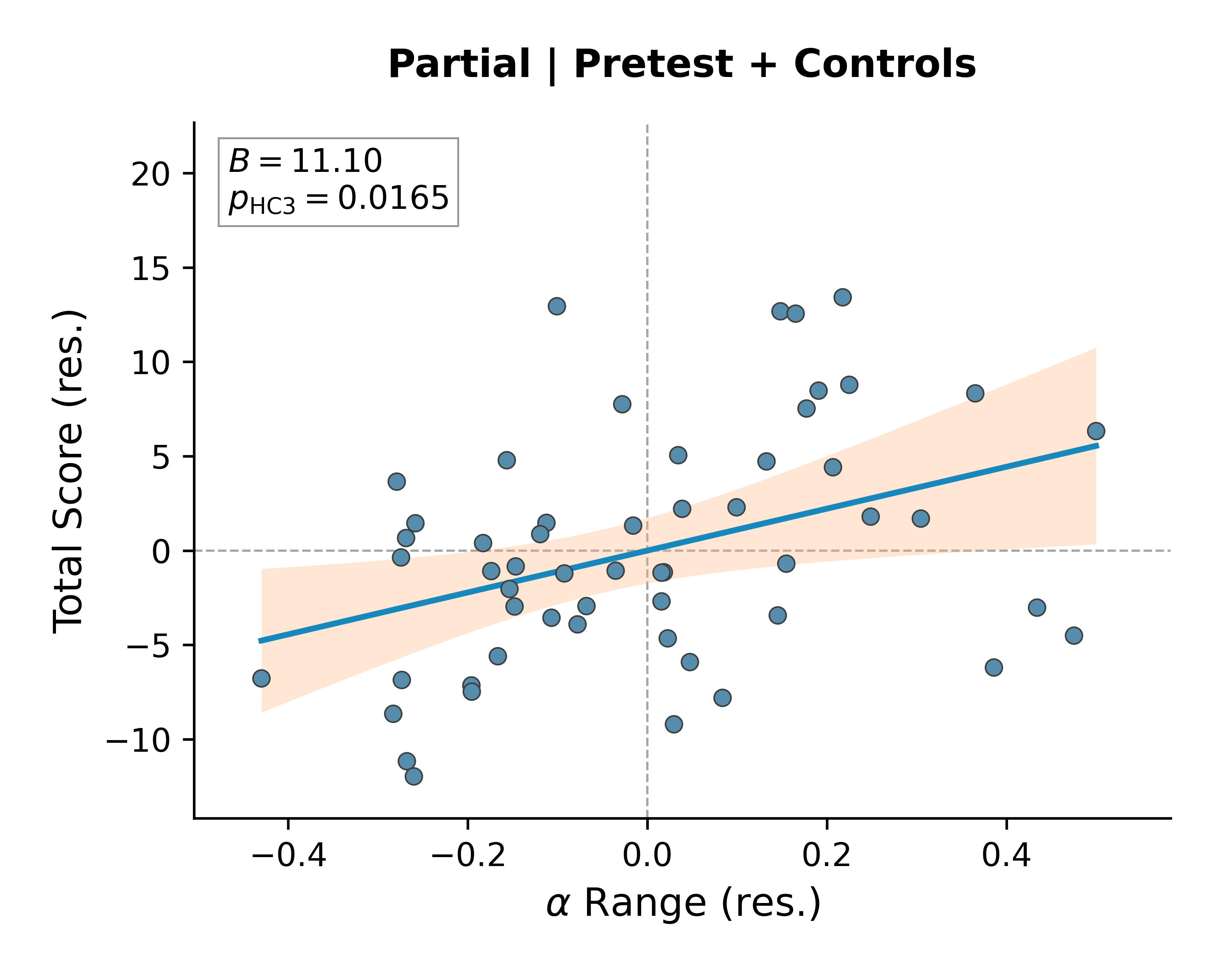}
  \end{minipage}
  \caption{\label{fig:flexibility}Between-task temporal variation and
  adjusted problem-solving performance ($N = 55$).
  (a)~Partial regression of Total Score on $\alpha_{\mathrm{range}}$
  after controlling for Pretest Performance (Model~1).
  (b)~Partial regression after controlling for Pretest Performance and
  four keystroke-derived covariates: mean characters produced, deletion
  rate, typing speed, and initial response latency (Model~3).
  Axes show residuals after regressing both $\alpha_{\mathrm{range}}$
  and Total Score on the respective covariates. Points represent students;
  lines show the fitted partial associations. Shaded bands indicate
  95\% confidence intervals for the fitted lines. Insets report
  unstandardized coefficients $B$ and HC3-robust $p$-values from the
  corresponding multiple-regression models. Relative to the same covariate model without $\alpha_{\mathrm{range}}$, Model 3 accounts for an additional 11.3 percentage points of variance in this sample ($\Delta R^2 = .113$).}
\end{figure}

To test its contribution beyond prior knowledge, we conducted a hierarchical
regression analysis ($N = 55$; Table~\ref{tab:regression}). The baseline model
including both Pretest Performance and $\alpha_{\mathrm{range}}$ explained
22.8\% of the variance in Total Score ($R^2 = .228$). The range of $\alpha$ values across tasks
was positively associated with performance after controlling for prior
knowledge ($B = 9.12$, 95\% CI $[1.27,\;16.98]$,
$p_{\mathrm{HC3}} = .0237$; Fig.~\ref{fig:flexibility}a).
The unadjusted association was also positive but did not reach statistical
significance ($r = .248$, $p = .068$).

A potential concern is that $\alpha_{\mathrm{range}}$ merely proxies general
behavioral differences such as producing more text, typing faster, or
revising more frequently. To address this, we estimated an extended model
including four keystroke-derived control variables: mean characters
produced, deletion rate, typing speed
(characters per minute), and initial response latency
(Table~\ref{tab:regression}, Model~3). All variance inflation factors were
below 1.65. In this model, $\alpha_{\mathrm{range}}$ remained positively
associated with Total Score ($B = 11.10$, 95\% CI $[2.12,\;20.07]$,
$p_{\mathrm{HC3}} = .0165$; standardized $\beta = .351$;
Fig.~\ref{fig:flexibility}b). Adding $\alpha_{\mathrm{range}}$ to a model already
containing Pretest Performance and all four controls increased explained
variance from $R^2 = .250$ to $.363$ ($\Delta R^2 = .113$), accounting for
an additional 11.3 percentage points of variance in this sample.
A 0.1-unit greater $\alpha_{\mathrm{range}}$ corresponded to an estimated
1.11-point higher Total Score at the same covariate values
(95\% CI $[0.21,\;2.01]$). As a sensitivity analysis, replacing $\alpha_{\mathrm{range}}$ with the
intra-individual standard deviation of $\alpha$ also yielded a positive
association in the fully controlled model ($B = 25.37$,
$p_{\mathrm{HC3}} = .0293$). 

To complement the tail-based analyses, we examined the locations
($\mu_j$), widths ($\sigma_j$), and weights ($w_j$) of the three fitted
mixture components across 320 participant--task fits from 55 students. After adjustment for pretest performance, no association
between participants' mean component parameters and total performance
reached HC3 $p<.05$ (smallest $p=.059$; smallest BH-adjusted $q=.472$
in a ten-test family including mean interval skewness). The estimated long-component location varied more across tasks within
participants than the shorter-component locations. Mean individual
across-task SDs were .040, .106, and .328 for $\mu_1$, $\mu_2$, and
$\mu_3$, respectively, on the $\log_{10}(t/\mathrm{s})$ scale
(both one-sided paired Wilcoxon comparisons of $\mu_3$ variability
with the shorter components: Holm-adjusted $p<.001$).
These SDs describe variation of fitted locations across tasks, distinct
from the within-component widths $\sigma_j$, and include estimation
uncertainty. Greater variability in $\mu_3$ was not significantly
associated with total performance after adjustment for pretest
performance (HC3 $p=.790$), nor did it add a significant association
beyond pretest performance and $\alpha_{\mathrm{range}}$
(HC3 $p=.664$). Thus, the performance association observed for
$\alpha_{\mathrm{range}}$ was not detected for this particular measure
of long-component variability. A separate analysis examined whether deviations in $\mu_3$ from a
student's own mean across tasks were related to that student's
task-standardized performance. In a mixed-effects model with participant
random intercepts, task fixed effects, and adjustment for within-person
differences in $\sigma_3$ and interval count, the estimated association
was negative ($B=-0.348$, nominal Wald $p=.022$), but did not survive
Benjamini--Hochberg correction across six component-location and
component-width coefficients ($q=.133$). The point estimate corresponds
to a 0.105-SD lower task score per doubling of the long-component median.
This within-person association therefore remains exploratory.

\begin{table}[t]
\caption{\label{tab:regression}Hierarchical regression summary for Total
Problem Solving Score ($N = 55$). Model~1 includes Pretest Performance and
$\alpha_{\mathrm{range}}$; Model~2 includes Pretest Performance and four
keystroke controls, without $\alpha_{\mathrm{range}}$; Model~3 adds
$\alpha_{\mathrm{range}}$ to Model~2. Covariate rows indicate model
inclusion; the reported coefficient and confidence interval refer to
$\alpha_{\mathrm{range}}$.}
\centering
\begin{tabular}{lccc}
\toprule
 & Model 1 & Model 2 & Model 3 \\
\midrule
Pretest Performance & Included & Included & Included \\
Mean Characters Produced & --- & Included & Included \\
Deletion Rate & --- & Included & Included \\
Typing Speed & --- & Included & Included \\
Initial Response Latency & --- & Included & Included \\
\midrule
$B_{\alpha_{\mathrm{range}}}$ & $9.12$ & --- & $11.10$ \\
95\% CI & $[1.27,\;16.98]$ & --- & $[2.12,\;20.07]$ \\
$p_{\mathrm{HC3}}$ & $.0237$ & --- & $.0165$ \\
\midrule
$R^2$ & $.228$ & $.250$ & $.363$ \\
$R^2_{\mathrm{adj}}$ & $.198$ & $.174$ & $.283$ \\
$\Delta R^2$ (M2 $\to$ M3) & --- & --- & $.113$ \\
$F_{\mathrm{change}}(1,48)$ & --- & --- & $8.48$ \\
$p$ for $F_{\mathrm{change}}$ & --- & --- & $.0054$ \\
\bottomrule
\end{tabular}

\begin{flushleft}
\footnotesize{$B$ = unstandardized coefficient. Coefficient confidence
intervals and $p$-values use HC3 heteroscedasticity-consistent standard
errors. The $F$-change test is conventional, not HC3-robust. All reported
$p$-values are nominal.}
\end{flushleft}
\end{table}

\section{Discussion}\label{sec12}

Complex cognitive tasks don't usually proceed as a series of continuous actions. Instead, they are organized into bursts of rapid production interleaved with pauses spanning multiple time scales. Using synchronized keystroke and eye-tracking data during multirepresentational problem solving, we show (i) a broad, heavy-tailed timing structure of IKIs within the observed range, (ii) a description in terms of overlapping temporal ranges, and (iii) systematic associations between pause duration and attention to external representations. The findings go beyond a simple description of timing structure in that we find a link between success and across-task variation in the fitted tail dynamics and that the performance relevance of these dynamics is contingent upon the task. Together, these results are consistent with adaptive reconfiguration of temporal organization across tasks.

One of the implications of this is with respect to the form of the timing variability. Pause durations are strongly right-skewed and exhibit a power-law-like decay over part of the observed range, with a substantially better tail fit than the exponential alternative. On the common analysis support, the lognormal alternative reached the power-law boundary and provided no likelihood improvement; the nominal AIC difference reflects the additional parameter rather than a likelihood advantage. At the same time, goodness-of-fit diagnostics did not support an exact power-law description. We therefore interpret the evidence as heavy-tailed timing within the analyzed window rather than exact scale invariance or a uniquely identified generative law. Functionally this is important as extended pauses are part of the behavioural distribution in complex goal directed tasks and so distributional descriptions complement the mean based summaries and fixed pause thresholds.

The mixture structure of the pause times also suggests that heavy tailed timing can be described in terms of overlapping temporal ranges. A Gaussian mixture model on log-pause duration provides a parsimonious three-component description: short and intermediate ranges with sub-second medians, and a long-pause range with broad dispersion. This common parameterization is favored by BIC over four components in most participant--task cells, although four components fit the pooled distribution better. It does not establish three discrete cognitive processes. While any mapping from timing alone to cognitive subprocesses must remain tentative, the temporal description gains functional context when aligned with gaze. Across the four derived duration categories, pooled valid dwell in the response field decreased from approximately 84\% to 23\%, while graph-directed dwell increased from 3\% to 51\% (Fig.~\ref{fig:gaze_fixations}). This pattern is consistent with local production monitoring during short pauses and greater external information uptake during very long pauses. Participant-level mean fixation durations also increased, although requiring fixations to fall entirely within a pause imposes duration-dependent selection. Together, these observations suggest that pauses are not merely ``gaps'' in typing, but can accompany changes in visual sampling and possible representational integration.

A second implication is one of individual differences and performance. The fitted tail exponent $\alpha$ provides a compact, model-dependent descriptor of the relative weighting of shorter versus longer intervals within the fitted tail. Although global $\alpha$ was not significantly associated with total score, a formal task-by-$\alpha$ interaction supported differences between task-specific associations ($\chi^2(5)$ = 19.93, p = .0013). Task~4 showed a positive association after FDR correction, whereas the negative estimate for Task~5 did not survive correction (Fig.~\ref{fig:corr}). This context dependence motivated examining within-person variation in tail dynamics across tasks, summarized by $\alpha_{\mathrm{range}}$. Greater $\alpha_{\mathrm{range}}$ was associated with higher overall performance after adjustment for prior knowledge and four keystroke covariates (B = 11.10, 95\% CI [2.12, 20.07], $\beta$ = .351, $p_{\mathrm{HC3}}$ = .0165; Fig.~\ref{fig:flexibility}, Table~\ref{tab:regression}). Adding $\alpha_{\mathrm{range}}$ to the control model accounted for an additional 11.3 percentage points of variance in this sample ($\Delta R^2$ = .113). This finding extends the performance account beyond measured differences in text production, deletion rate, typing speed, and initial response latency, without establishing out-of-sample predictive validity.

Complementary component analyses further refine this interpretation. The fitted long-component location, $\mu_3$, varied more across tasks within participants than the shorter-component locations, but this variability was not significantly associated with total performance, either after adjustment for pretest or after additional adjustment for $\alpha_{\mathrm{range}}$. The average component profiles also showed no significant pretest-adjusted associations. The current results therefore do not support treating greater temporal variability of any kind as uniformly beneficial. Within persons, positive deviations in $\mu_3$ from a student's own across-task mean were nominally associated with lower task-standardized performance, but this relationship did not survive multiplicity correction (q = .133). This exploratory pattern concerns differences between tasks within a person, not the effect of prolonging an individual pause. 

In the light of the adaptive expertise framework this pattern is consistent with the view that successful performance on non-routine problems requires a balance between efficiency and innovation \cite{schwartz2000efficiency,hatano1984two}. Our data do not establish a single task-independent association between fitted tail dynamics and success. Instead, the task-dependent $\alpha$--performance associations and the positive adjusted association between $\alpha_{\mathrm{range}}$ and overall performance suggest that successful solvers may reorganize temporal dynamics as task demands change. The range measures the extent to which fitted tail dynamics vary across tasks, rather than whether those variations are appropriate to the task at hand. We therefore use temporal adaptivity as a theoretical interpretation rather than a validated ability measured by $\alpha_{\mathrm{range}}$. Together with the gaze results, this pattern is compatible with relatively fluent local production interleaved with episodes of external information uptake and possible representational integration. The gaze analyses provide a behavioral context for this interpretation, without establishing which perceptual processes explain the performance association.

The present study is in the domain of physics problem solving, a domain which involves representational coordination between text, diagrams and symbolic relations. We stress however that the core temporal phenomena need not be specific to that domain. Heavy-tailed timing, multiple overlapping temporal ranges, and the association of longer pauses with perceptual sampling may also be relevant when people coordinate internal production with external representations under changing subgoals---such as mathematical proof construction, programming, design sketching, or analytical writing. Physics tasks here function as a testbed for this broader class of complex cognitive activities; whether the same performance associations generalize to other activities remains to be tested.

Our results also connect to a wider literature on scaling in cognition and behavior
\cite{WangEtAl2011,PalvaEtAl2013,JonesEtAl2023}. Heavy-tailed interval distributions and fractal temporal correlations are distinct properties: the distribution of pause durations does not capture their sequential ordering. While the present data do not permit inference about neural mechanisms or criticality, combining a distributional timing description with gaze associations provides a descriptive bridge between micro-level production intervals and attention to external representations. The contribution is therefore a characterization of task-related temporal organization, rather than evidence for a scale-free control mechanism.

A practical takeaway is that keystroke timing offers a low-barrier window into the temporal organization of cognitive activity. Unlike eye tracking, keystroke logs can be collected unobtrusively and at scale in everyday digital environments. This opens two applied directions for future validation. First, process-aware adaptive systems could respond not only to correctness but to changes in the temporal organization of activity---for example, by timing feedback or prompts to support productive coordination of production and information integration. The present $\alpha_{\mathrm{range}}$ measure summarizes completed tasks and has not been validated as an online support trigger. Second, the same traces can be integrated with sequential and process-mining approaches to model transitions between production, inspection of external representations, and revision. In human--AI interaction settings, such traces could also help characterize how tool use reshapes temporal organization; this remains a prospective research direction rather than an application established here.

There are several limitations which constrain the generalisability of the findings. The sample consists of 55 university students and six problems from a single task family; replication across cohorts, task families and domains is required to determine the boundary conditions. The performance analyses are correlational and cannot establish whether temporal variation reflects a stable trait, a learned strategy, or an emergent response to task structure. The controlled models estimate an association with the observed range; they do not separate genuine temporal variation from uncertainty in the fitted exponents. Range estimates can also depend on fitted cutoffs and the number of available tasks. The 50-ms--100-s analysis window and the use of sampled text-content changes rather than directly recorded key-down events further bound the temporal interpretation. Finally future work should look at richer hierarchical or state space models which treat task context, individual differences and parameter uncertainty jointly and assess whether experimentally varied task demands lead to appropriate temporal adjustments and whether support for such adjustments leads to improved performance.

Overall the findings suggest that understanding what constitutes successful performance in problem solving requires an attention to task context and temporal flexibility rather than simply assuming that there is some one 'optimal' timing profile. The adjusted $\alpha_{\mathrm{range}}$ association and the gaze-based characterization of pauses together motivate adaptive temporal reconfiguration as a hypothesis for further process-level investigation.

\section{Conclusion}\label{sec13}

Our integration of keystroke and gaze data shows that when students are solving representation-rich tasks, their work is characterized by a multi-timescale rhythm of broad, heavy-tailed pause dynamics which can be described by in terms of overlapping temporal ranges and are linked to systematic differences in visual attention. The timing--performance associations vary across tasks and the greater the across-task variation in fitted tail exponents the higher the overall performance after adjustment for prior knowledge and measured keystroke covariates. This supports a temporal flexibility perspective on successful problem solving and is in line with the notion of adaptive reconfiguration of temporal organisation. Fine-grained, unobtrusive process data may therefore be of value as a complement to product based assessment and may be used to inform process aware educational technologies. Their use to guide support for individual learners will of course require independent validation and intervention studies.

\section{Methods}\label{sec11}

\subsection{\label{sec:subsection-5}Participants and Procedure}
The study involved 60 university students in the early stages of
their degree programs. RParticipants were recruited via course mailing lists and via announcements made in lectures. Participation was entirely voluntary and participants were compensated for their time. All participants provided written informed consent prior to data collection.  Data from five participants were excluded because eye-tracking
calibration was unsuccessful ($n=3$) or the tasks were not fully
completed ($n=2$). The final analytic sample therefore comprised
55 participants. At the start of each session, the researcher explained the study purpose, the procedure,
and the basics of eye tracking. Each participant was seated at a laptop connected to a
24-inch external monitor equipped with a Tobii Pro Nano eye tracker \cite{TobiiProSpark}.
The eye tracker recorded gaze data at 60\,Hz. The manufacturer reports an accuracy of
0.3$^\circ$ and a precision of 0.10$^\circ$ RMS under optimal
conditions, a latency of one frame (17\,ms), and an operating distance of 45--85\,cm.
Participants were seated within this range from the tracker in a controlled indoor
environment with uniform lighting.

A 5 point calibration was completed for each participant using Tobii Pro Eye Tracker Manager prior to each session and the quality of the calibration was checked by the experimenter. Participants were instructed to keep their heads as still as possible and no recalibration was carried out during the session. All of the materials and responses were in German. Participants first completed 17 single-choice items (see Fig.~\ref{fig:singlechoice})
assessing prior knowledge, followed by six open-ended problems from introductory
electromagnetism/electrostatics (see Fig.~\ref{fig:openquestion}). The open-ended items
combined textual prompts with schematic diagrams (multiple representations) and
required free-form written explanations in full sentences (e.g., charged particle
motion in electric/magnetic fields, induction in coupled loops, and interactions of
point charges). For the open-ended questions we recorded gaze data and changes to the response input field on a shared timeline. On completion the response files were saved and all the process data exported for analysis. The eye tracking data were processed to produce the standard gaze measures and fixations were detected using an I-VT (velocity threshold) algorithm \cite{Salvucci2000}.

\subsection{\label{sec:subsection-6}Task Environment}
The task environment was implemented as a custom desktop application built with Electron
(HTML/CSS/JavaScript, Chromium-based renderer). The application presented the study in a
slide-based, forward-only flow: participants advanced via a \emph{Continue} button and could not
return to previous slides. This design was chosen to standardize exposure and reduce uncontrolled
re-reading behavior.

Eye movements were streamed from the Tobii software stack into the app via a WebSocket. Each gaze sample was time-stamped upon reception and mapped to the currently displayed
slide and predefined Areas of Interest (AOIs; problem text, graph/diagram region, response field,
instructions, background). AOIs were defined a priori for each open-response slide based on the
static screen layout. To enable a frame-aligned analysis of the gaze and writing data we sampled the response input field at the same rate as the eye tracking stream (60\,Hz). We defined a keystroke-equivalent input-update
event whenever the input-field content changed between two consecutive samples and computed
inter-event intervals at 16.7\,ms temporal resolution. As a validation of event granularity, we quantified the distribution of update sizes ($\Delta$ of characters per sample): 99.7\% of events changed the response by exactly one character (insertions and deletions), whereas multi-character updates were rare
(0.286\%) and excluded in sensitivity analyses. This allowed for a frame-aligned analysis of the within-pause gaze behaviour on the same samples as the writing trace.

\subsection{\label{sec:scoring}Scoring of Open-Ended Problems}

To evaluate participants' performance on the six open-ended problems,
a detailed coding scheme was developed, adapted from the
problem-solving rubric proposed by Docktor et al.~\cite{DocktorEtAl2016}.
For each problem, the scheme defined multiple specific sub-criteria.
The final performance score for a task was derived by summing
the scores of these individual criteria. Two independent raters scored all responses based on this scheme. Because the rating process was conducted at the level of specific binary sub-criteria, inter-rater reliability was assessed using Cohen's Kappa for each criterion. The values reported in Table~\ref{tab:kappa} represent the mean Cohen's Kappa across all sub-criteria within each respective task. As shown, the reliability was consistently high across all tasks, ranging from mean $\kappa = .72$ to $\kappa = .91$, indicating substantial to almost perfect agreement \cite{Landis1977}.

\subsection{\label{sec:subsection-7}Data Analysis}

At least 50 IKIs were required for a participant- task fit in order to estimate the power-law (with unbounded upper support) and 100 IKIs for the GMMs.
Power-law cutoffs were selected separately for each participant–task combination via KS minimisation and the exponent was then estimated from the IKIs at or above the selected cutoff.
GMMs were fitted to $\log_{10}(t/\mathrm{s})$ with covariance
regularization of $0.001$ and components ordered by their means;
the stable solution with the highest log likelihood across initialisations was kept.

\subsubsection{Across-task variability in tail exponents}

To look at the variability in the students' writing processes we calculated the range of the fitted tail exponents across tasks.
This quantity summarizes the intra-individual variability (IIV)
of $\alpha$ across the six problem-solving tasks.
Task-specific $\alpha$ estimates were upper-winsorized at 3.
The range, $\alpha_{\text{range}}$, was calculated as the difference
between each participant's highest and lowest values across all
available valid tasks:
\begin{equation}
    \alpha_{\text{range}} = \max(\alpha_{1\dots6}) - \min(\alpha_{1\dots6})
\end{equation}
This is a measure of the extent to which a student's fitted tail dynamics vary across tasks. Sensitivity analyses were carried out to check the robustness of the observed performance associations.

\clearpage

\section*{Declarations}

\subsection*{Funding}

This research was funded by the German Research Foundation
(Deutsche Forschungsgemeinschaft, DFG; CORE, project number
462702138, subprojects A03 and C07; and "Real-world anchored
augmented visualizations to support conceptual learning in
physics", project number 517713394) and supported by
LMUexcellent, funded by the Federal Ministry of Education
and Research (BMBF; now the Federal Ministry of Research,
Technology and Space, BMFTR) and the Free State of Bavaria
under the Excellence Strategy of the Federal Government
and the Länder.

We also gratefully acknowledge the support of subproject A05
of the DFG-funded Collaborative Research Centre SHARP
(TRR 419, project number 542251251).

J.~Z. was funded by the Deutsche Forschungsgemeinschaft (DFG)
via SFB 1690/1, A07.

\subsection*{Competing interests}

The authors declare no competing interests.

\subsection*{Ethics approval and consent to participate}

The study was approved by the Ethics Committee of
LMU (approval number EK-MIS-2024-241).
All participants provided written informed consent
prior to participation. Participation was voluntary
and compensated.

\subsection*{Data availability}

The processed and de-identified inter-keystroke interval
(IKI) data supporting the analyses of this study will
be made publicly available in a repository upon publication.

The underlying data can be made available to editors
and reviewers during peer review, subject to applicable
data protection requirements.

Raw eye-tracking and keystroke-level data are not
publicly available due to participant privacy and
ethical restrictions.

\subsection*{Code availability}

The custom Python code used for data processing,
statistical analyses, and figure generation is
available from the corresponding author to editors
and reviewers upon request.

\subsection*{Author contributions}

\textbf{Conceptualization:} Y.D., F.K., P.W., J.K., S.K.;
\textbf{Methodology:} Y.D., F.K., P.W., J.K., S.K.;
\textbf{Investigation:} Y.D., F.K.;
\textbf{Software:} Y.D.;
\textbf{Data curation:} Y.D., F.K., N.S., S.K.;
\textbf{Formal analysis:} Y.D., J.Z., N.S., K.A., S.K.;
\textbf{Visualization:} Y.D., J.Z., K.A., S.K.;
\textbf{Writing -- original draft:} Y.D.;
\textbf{Writing -- review \& editing:} Y.D., F.K., P.W., J.Z., N.S., K.A., J.K, S.K.;
\textbf{Supervision:} J.K., S.K.;
\textbf{Funding acquisition:} J.K., S.K.

All authors reviewed and approved the final manuscript.

\clearpage

\begin{appendices}
\setlength{\intextsep}{10pt}

\FloatBarrier

\section{Additional tables and figures}
\label{secA1}

\setlength{\fboxrule}{0.1pt}
\setlength{\fboxsep}{5pt}

\begin{figure}[H]
  \centering
  \fbox{\includegraphics[width=\dimexpr\linewidth-2\fboxsep-2\fboxrule\relax]{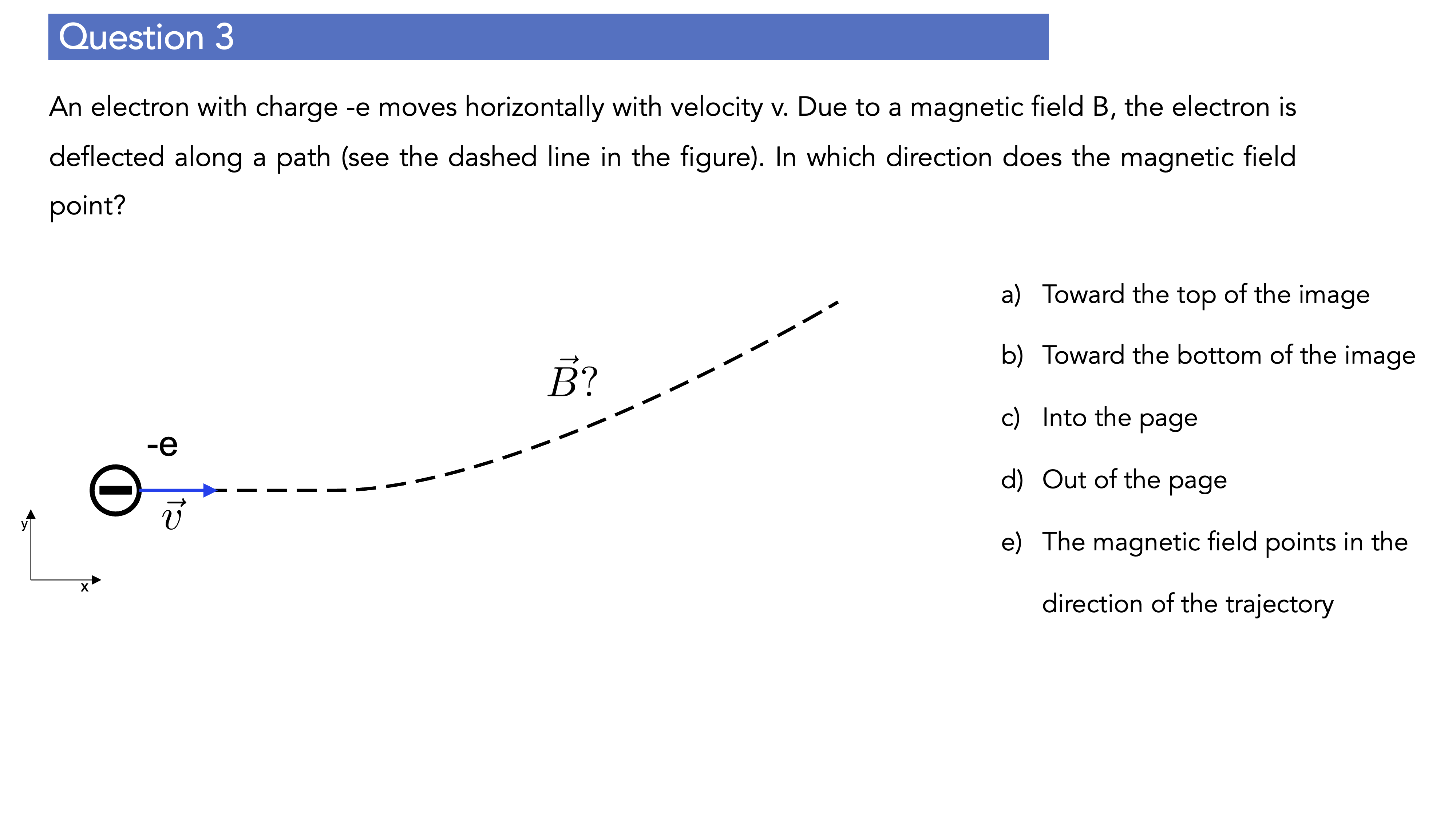}}
  \caption{\label{fig:singlechoice} Example of a single-choice item, translated from German.}
\end{figure}

\begin{figure}[H]
  \centering
  \fbox{\includegraphics[width=\dimexpr\linewidth-2\fboxsep-2\fboxrule\relax]{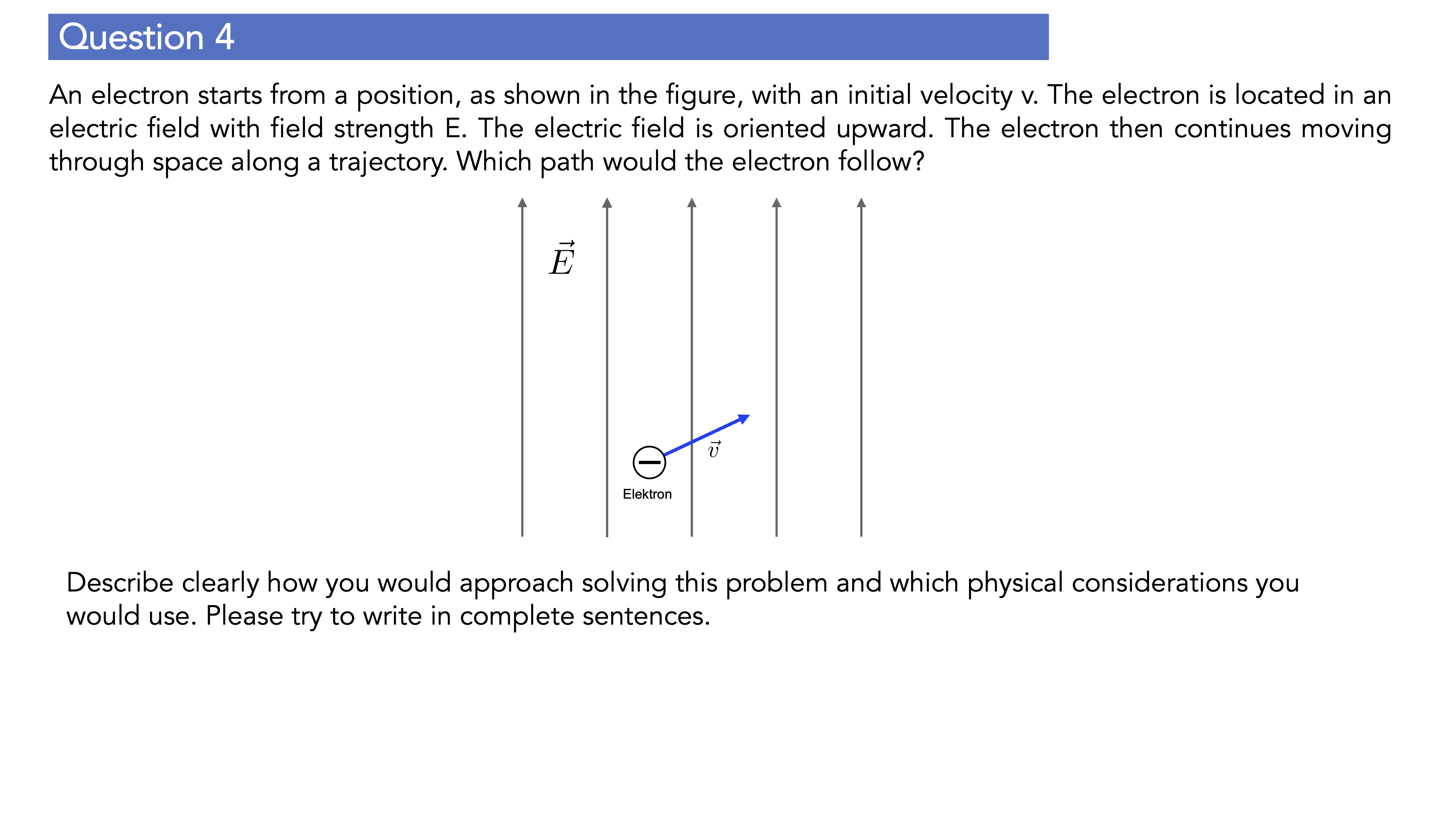}}
  \caption{\label{fig:openquestion} Example of an open-ended problem, translated from German.}
\end{figure}

\clearpage
\noindent
\begin{minipage}[t]{0.70\textwidth}
\vspace{0pt}\small
\captionsetup{hypcap=false,justification=raggedright,singlelinecheck=false}
\centering
\captionof{table}{Three-component GMM on $\log_{10}(t_{\mathrm{pause}}/\mathrm{s})$
for IKIs between 50\,ms and 100\,s. For each component we report the weight,
the mean and standard deviation on the $\log_{10}$ scale, and the median on
the original time scale ($10^{\mu}$).}
\label{tab:gmm-params}
\begin{tabular}{S[table-format=1.0]
                S[table-format=1.3]
                S[table-format=+1.3]
                S[table-format=1.3]
                S[table-format=1.3]}
\toprule
{Component} & {Weight} & {Mean ($\log_{10}$\,s)} &
{Sigma ($\log_{10}$\,s)} & {Median (s)} \\
\midrule
1 & 0.761 & -0.799 & 0.185 & 0.159 \\
2 & 0.186 & -0.322 & 0.244 & 0.476 \\
3 & 0.052 & +0.228 & 0.510 & 1.690 \\
\bottomrule
\end{tabular}
\end{minipage}\hfill
\begin{minipage}[t]{0.27\textwidth}
\vspace{0pt}\small
\captionsetup{hypcap=false,justification=raggedright,singlelinecheck=false}
\centering

\centering
\captionof{table}{Averaged Cohen's $\kappa$ across individual criteria for the six open-ended problems.}
\label{tab:kappa}
\begin{tabular}{l S[table-format=1.3]}
\toprule
Task & {Mean $\kappa$} \\
\midrule
Problem 1 & 0.763 \\
Problem 2 & 0.888 \\
Problem 3 & 0.907 \\
Problem 4 & 0.835 \\
Problem 5 & 0.792 \\
Problem 6 & 0.722 \\
\bottomrule
\end{tabular}

\end{minipage}

\begin{figure}[H]
  \centering
  \includegraphics[width=0.92\linewidth]{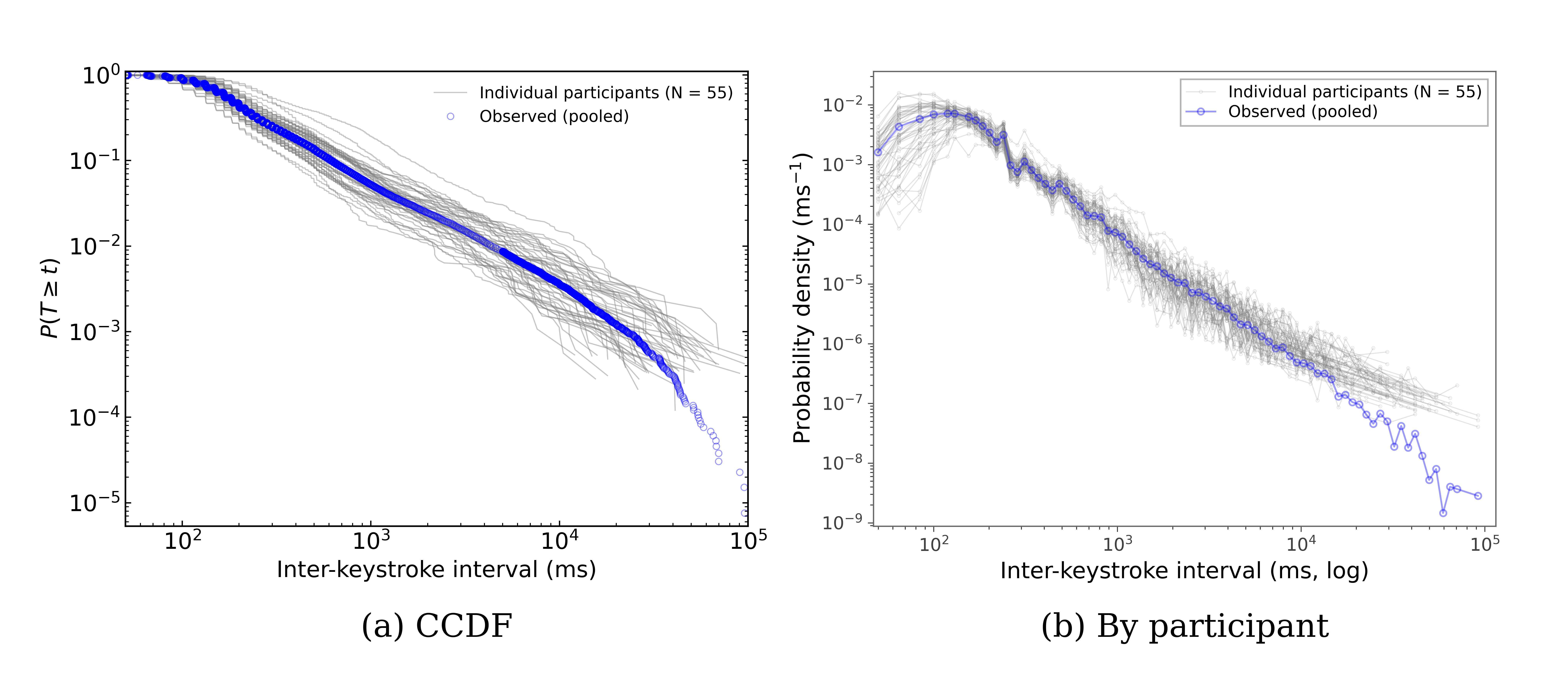}
\caption{\label{fig:supp-participant-distributions}
Participant-level IKI distributions ($N=55$).
(a) Empirical CCDFs. (b) Normalized histogram densities.
Gray curves represent individual participants across their available
tasks; blue shows all 131,718 IKIs pooled across participants
(50\,ms--100\,s).}
\end{figure}

\begin{figure}[H]
  \centering
  \includegraphics[width=0.7\linewidth]{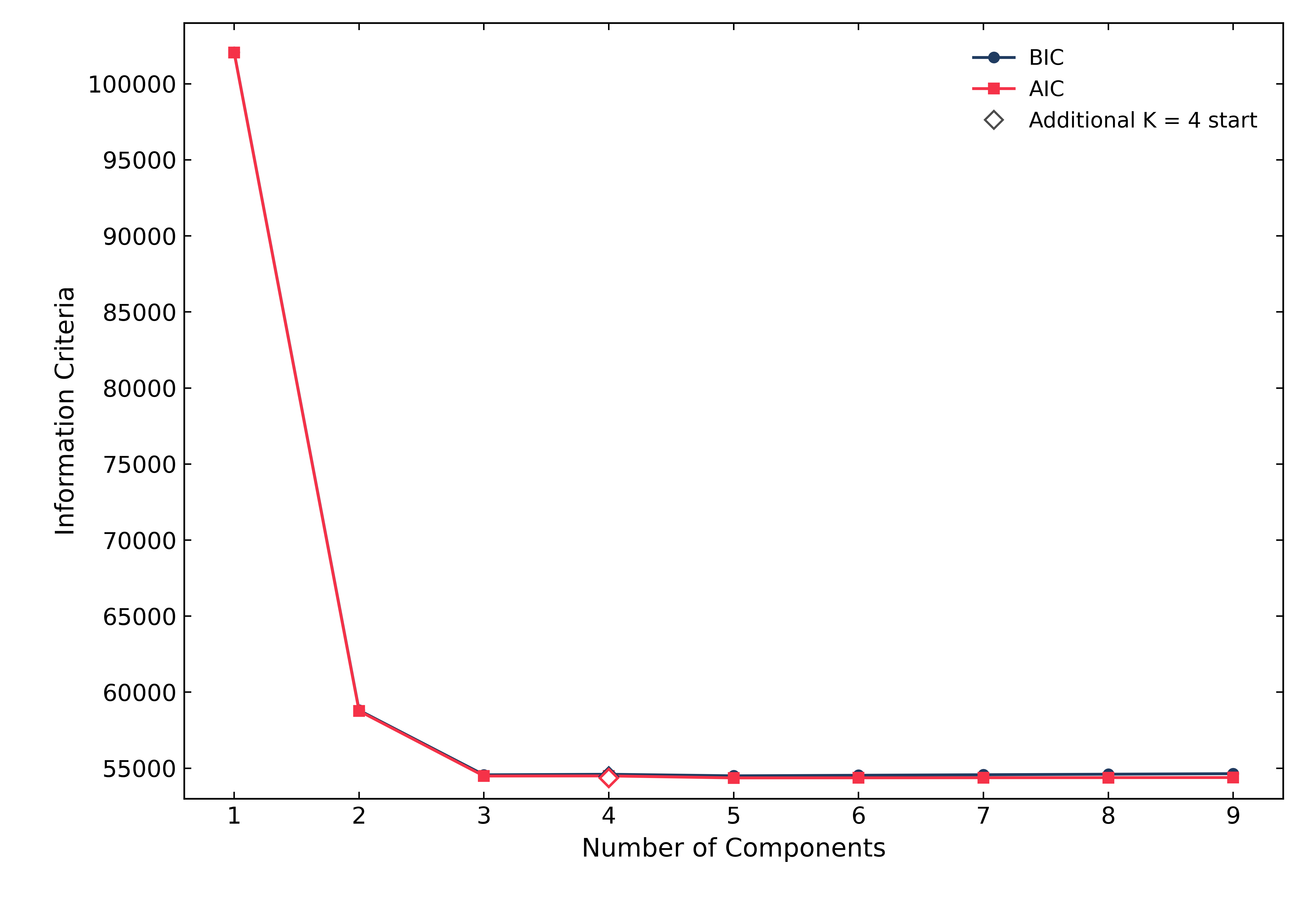}
\caption{\label{fig:gmm_bic_aic}
Gaussian-mixture model comparison for log-transformed
IKIs. We retained $K=3$ as a parsimonious,
interpretable description of short, intermediate, and long intervals.
The lowest observed AIC/BIC occurred at $K=4$
(open diamonds: additional initialization).}\end{figure}

\FloatBarrier

\clearpage

\section{Tail diagnostics and robustness checks}
\label{app:tail-diagnostics}

For the unbounded reference fit, the descriptive $\log_{10}$-CCDF
regression yielded a slope of $-1.2432$ ($R^2=.994$), compared with
$-(\hat{\alpha}-1)=-1.2692$; the two-sided KS distance was $D=.02382$.
This regression is a graphical summary, not a goodness-of-fit test.

Conditioning the candidate densities on
$366.183$\,ms $\leq x\leq100$\,s, with the lower cutoff held fixed,
yielded $\hat{\alpha}=2.2616$, compared with $\hat{\alpha}=2.2692$
for the unbounded reference fit. The lognormal fit provided no gain
in maximized log likelihood; its nominal $\Delta\mathrm{AIC}=2$
reflects the additional nominal parameter
(Table~\ref{tab:tail-compare}). For the shared conditioned
power-law/lognormal limiting density, parametric-bootstrap KS
diagnostics yielded $D=.02481$ and Monte Carlo $p=.0005$ for each
model (1,999 simulations each).

\begin{table}[htbp]
\centering
\caption{Tail model comparisons on $26{,}168$ IKIs, with all densities
conditioned on $366.183$\,ms $\leq x\leq100$\,s
($\hat{\alpha}=2.2616$ for the power law).
$\Delta\ell=\ell_{\mathrm{alt}}-\ell_{\mathrm{PL}}$ and
$\Delta\mathrm{AIC}=\mathrm{AIC}_{\mathrm{alt}}-\mathrm{AIC}_{\mathrm{PL}}$;
positive values indicate a higher log likelihood or a higher nominal
AIC, respectively, for the alternative.}
\label{tab:tail-compare}
\centering
\begin{tabular}{lrr}
\toprule
Alternative & $\Delta\ell$ & $\Delta\mathrm{AIC}$ \\
\midrule
Lognormal$^{a}$ & $0$ & $2.00$ \\
Exponential & $-12{,}759.026$ & $25{,}518.05$ \\
Stretched exponential$^{a}$ & $\approx0$ & $2.00$ \\
Truncated power law$^{a}$ & $\approx0$ & $2.00$ \\
\bottomrule
\end{tabular}

\begin{flushleft}
\footnotesize
$\ell$ denotes the log-likelihood (natural logarithms).
$^{a}$The optimum reaches or approaches the power-law boundary.
The $+2.00$ AIC differences arise solely from the additional nominal
parameter at the limiting fit; they are neither significance tests
nor model probabilities. ``Truncated power law'' denotes an
exponential cutoff, in addition to the common 100-s conditioning.
\end{flushleft}
\end{table}

The lognormal curvature optimum lay at the power-law boundary,
with no likelihood gain. Predictive log scores were identical in all
55 leave-one-participant-out folds, and the lognormal fit reached
the power-law boundary in 4,980 of 5,000 whole-participant bootstrap
samples (99.6\%). These comparisons are conditional on the selected
analysis window.

For consecutive single-character insertions ($n=113{,}474$),
the unbounded reference fit yielded $\hat{\alpha}=2.274$ at
$x_{\min}=366.183$\,ms, with $n_{\mathrm{tail}}=17{,}954$.
In the full analysis, the aggregated participant-level and task-level
exponent ranges were $[1.95,2.86]$ and $[2.20,2.29]$, respectively.

\end{appendices}

\clearpage   
\FloatBarrier

\bibliographystyle{unsrtnat}
\bibliography{references}

\end{document}